# A Machine Learning Approach To Prevent Malicious Calls Over Telephony Networks

Huichen Li[1]    Xiaojun Xu[1]    Chang Liu[2]    Teng Ren[3]    Kun Wu[3]
Xuezhi Cao[1]    Weinan Zhang[1]    Yong Yu[1]    Dawn Song[2]
[1]*Shanghai Jiao Tong University*    [2]*University of California, Berkeley*    [3]*TouchPal Inc.*

*Abstract*—Malicious calls, i.e., telephony spams and scams, have been a long-standing challenging issue that causes billions of dollars of annual financial loss worldwide. This work presents the first machine learning-based solution without relying on any particular assumptions on the underlying telephony network infrastructures.

The main challenge of this decade-long problem is that it is unclear how to construct effective features without the access to the telephony networks' infrastructures. We solve this problem by combining several innovations. We first develop a TouchPal user interface on top of a mobile App to allow users tagging malicious calls. This allows us to maintain a large-scale call log database. We then conduct a measurement study over three months of call logs, including 9 billion records. We design 29 features based on the results, so that machine learning algorithms can be used to predict malicious calls. We extensively evaluate different state-of-the-art machine learning approaches using the proposed features, and the results show that the best approach can reduce up to 90% unblocked malicious calls while maintaining a precision over 99.99% on the benign call traffic. The results also show the models are efficient to implement without incurring a significant latency overhead. We also conduct ablation analysis, which reveals that using 10 out of the 29 features can reach a performance comparable to using all features.

## I. INTRODUCTION

Spams and scams through telephony networks have caused an annual financial loss that is worth billions of dollars all over the world [1]–[3]. We refer to them as *malicious calls*. Unfortunately, there has been no simple and effective solution to stop them [33]. Although countries, such as US, have established National Do Not Call Registry to mitigate the issue, the problem is more severe in countries such as China, where such legislation is not available.

One of the main challenge is the lack of effective information for accurate malicious calls detection. Different from traditional email spams, malicious calls typically demand instant responses before the content in the call has been heard. Thus, only the header information can be used to prevent malicious calls from causing recipients to lose time, money, and productivity. Prior malicious call prevention techniques mainly focus on Spam over Internet Telephony (SPIT), and rely on server side information about the caller to predict malicious calls [7], [11], [19], [21], [24], [31], [34], [36], [40]. However, such information is typically unavailable for the end users on traditional telephony networks.

In this work, we focus on the malicious call prevention problem without relying on any particular underlying telephony network infrastructure. Thus the first challenge is how to gather effective information. The first contribution of this work is to collect information about malicious callers in order to build an effective prevention mechanism, using the TouchPal user interface. The basic idea is to implement TouchPal as a functionality of a mobile App that has a large number of users. Then TouchPal allows its users to label a finished call as malicious or not, and implements a simple reputation-based black-listing prevention mechanism based on users' tagging. TouchPal also promptly suggests users to label suspicious calls. This design also increases the tagged call log volume in addition to users' voluntary labeling through malicious call reporting services, such as 12321. In doing so, we can gather a large call log dataset without relying on any particular telephony network infrastructure.

Although the simple black-listing approach is effective, it has to observe enough call records from one malicious number before TouchPal can black-list the number. Our next question is: *how can we build an effective mechanism to detect a malicious call number early without answering too many calls dialed from the number?* We seek a machine learning solution.

The main obstacle is to design a set of effective *features*. To this end, we rely on the data logs collected from TouchPal to analyze which information is effective to be used as a feature. In fact, over the past several years, TouchPal has reached over 56 million daily active users and kept track of billions of call records monthly. Nowadays, TouchPal maintains the largest call log databases in China with respect to both call ID volume and call tag volume.

Using this dataset, we conduct a large scale measurement study to understand which information is more helpful to distinguish malicious calls from benign ones. Note that there have been several prior work providing measurement studies [15], [22]. However, these work mainly focused on malicious call records, and did not provide insights on how malicious call records differ from benign ones. Also, they focused on US call records, and it is unclear whether the same conclusions apply to Chinese malicious call ecosystems. MobiPot [9] provides a study to overcome these two issues; however, the study relies on only less than 700 call records, and we show that the observations from [9] are not robust when we increase the samples size by 7 orders of magnitude.

Our study overcomes all these issues, and sheds new light on the feature design, which is the core problem of this work. Our results reveal that (1) the provincial malicious call volume is more sensitive to the province's Gross Domestic Product

(GDP) than benign calls; (2) malicious calls are more likely to happen in a workday and during working hours than benign calls; and (3) the volume of incoming and outgoing calls from a number is indicative to distinguish malicious calls from benign calls. To the best of our knowledge, we are the first to present these findings with respect to distinguishing malicious calls from benign ones.

Inspired by our measurement study, we design 29 features for the malicious call prediction problem to include not only static information about the current call, but also extended information by examining *historic records* about the caller of an incoming call and by *cross-referencing* multiple records. We extensively evaluate the effectiveness of these features by using several state-of-the-art models. To this end, we use both the standard AUC score, and design a new metric, *average first prediction* (AFP). AFP is designed to evaluate the averaged amount of malicious calls that needs be observed before an approach can predict it as a malicious caller, without affecting benign call traffics. Our evaluation shows that using our proposed features, a random forest model can achieve an AUC score of at least **0.99**; further, it reduces the averaged necessary observed malicious calls by up to **90%** from a black-listing approach, while guaranteeing that over **99.99%** of the benign calls will not be blocked. In other words, the best random forest model using our 29 proposed features can reduce **90%** *unblocked* malicious calls.

Also, the evaluation shows that a neural network model can achieve a similar accuracy performance as the best random forest, but incurs a low latency overhead of less than 1ms. This shows that the models in our evaluation can be efficiently implemented on top of the current infrastructure to achieve both high accuracy and high efficiency.

We further conduct ablation study to understand the effectiveness of each proposed feature, and our evaluation shows that only 10 features are necessary to reach a high accuracy instead of the entire 29 features.

We summarize our contributions as follows.

1) We develop the TouchPal user interface to keep track of malicious calls and benign calls. Using this approach, TouchPal has maintained the largest call log database in China with respect to call ID volume, total call record volume, and malicious call volume;
2) We conduct a measurement study on the large scale call logs without sensitive user information to draw insights to design effective features for a machine learning-based malicious call prevention approach;
3) We propose 29 features, and extensively evaluate 6 state-of-the-art machine learning approaches. The results show that the best random forest model can achieve an AUC score of at least 0.99, and reduces up to 90% unblocked malicious calls compared with a black-listing approach, while at least 99.99% of the benign traffic will not be blocked;
4) We evaluate the model's runtime performance, and show that some of the performant models incur small latency overhead. Thus, the proposed approach can be efficiently implemented on top of the current structure;
5) To further understand the effectiveness of the proposed features, we conduct ablation analysis. We find that some features are more useful than others, and in an extreme case, using the top-10 most useful features can achieve a comparable performance to using all 29 features.

## II. OVERVIEW

In this section, we present an overview of the malicious call prevention problem and the TouchPal solution. We will first briefly review the malicious call status in China, and then define the problem by providing the requirements of malicious call prevention. We will then give an overview of the TouchPal solution, with the highlights of our technical development in this paper.

### A. Malicious calls in China

The legislation status in China against malicious calls is pre-mature. Services, such as National Do Not Call Registry in US, have not been available in China. The main channel provided by Chinese government is 12321, a dedicated service for reporting malicious calls and SMS messages. However, 12321 mainly relies on users' volunteer reporting; also, it is unclear how this information is eventually used.

Two policies are enforced by Chinese government, which may have effect on malicious calls. First, telecommunication providers in China are required to register a real-identity with every phone number. Second, a number dialing another in a different province will incur a long-distance cost. These two policies may increase the cost for a malicious caller. However, starting from September 1st, 2017, the long-distance fee has been canceled along with roaming cost [4].

### B. Malicious call prevention

We consider the problem to prevent malicious calls on the mobile side. That is, a malicious call preventer is implemented on the mobile phones to provide the service to detect whether an incoming call will be a malicious call. As we will explain in Section III, we mainly consider harassing and phishing calls as malicious calls. We have the following requirements.

**Without the access to the underlying telephony network infrastructure.** We require the solution to be deployed on end-users' devices; so, it does not have access to many information about the caller that is only available from the servers in telecommunication providers. This eliminates most of the existing SPIT prevention proposals. However, we emphasize that this requirement does not prevent a solution leveraging an server to collect and store information reported from the mobile devices.

**Light-weight for users.** The prevention mechanism should not incur many additional operations to end users. Ideally, the users should receive a benign call or dial a number as usual, and only need to operate the phone differently when the incoming call is predicted as a malicious call.

**Effectiveness.** The solution should not prevent users from receiving benign calls. A majority of the benign calls (i.e., $\geq 99.99\%$) should pass through the malicious call detector.

**Early detection.** Ideally, an effective malicious call preventer should start blocking all malicious calls from a phone number when as few malicious calls have been made as possible.

**Efficiency.** The malicious call preventer should incur a low latency overhead on the phone side to detect whether an incoming call is a malicious call. Ideally, the latency overhead should be 10ms or lower.

*C. Solution overview*

We now present the overview of our solution to use machine learning for malicious call prevention. We built TouchPal (Section III) as an additional functionality on top of a mobile App, which has hundreds of millions of users. TouchPal provides the functionality to allow users to label a phone call as malicious, and employs a reputation-based black-listing malicious call prevention mechanism.

However, the black-listing approach requires the same number to be labeled multiple times before it can be marked as a malicious caller and blocked. To mitigate this issue, we develop a machine learning approach to predict whether a phone number is a malicious caller before it has made too many malicious calls.

The main challenge is how to design effective features without tapping users call content. To this end, TouchPal keeps a call log containing each call record about a TouchPal user. In the call record, only less sensitive information such as call duration and call time is stored.

Though the call log hides sensitive information, its scale allows us to make important observations. We conduct a large scale measurement study (Section IV) using data over a period of three months containing 9 billion call records, and examine which information is more helpful to distinguish malicious callers from benign ones. Our study thus sheds light on the design of the selection of features (Section V).

Intuitively, besides the basic set of features about the current call, information from historic call records and information from multiple records can be useful in detecting malicious calls. We extensively evaluate several state-of-the-art machine learning approaches (Section VI), including random forest, neural networks, SVM, and logistic regression, and we observe that most of the models can achieve a high performance. First, the AUC scores of most models can achieve $0.99$ or higher. More importantly, we enforce the precision of benign calls to be at least $0.99$, and evaluate how many malicious calls need be observe before the number can be detected. Our evaluation shows that a random forest or a neural network only need to observe 2.5 calls on average to detect a malicious caller; we thus reduce up to $90\%$ unblocked malicious calls using the current black-listing approach in TouchPal.

### III. TouchPal FOR MALICIOUS CALL PREVENTION

In this section, we explain the pipeline of TouchPal to help the users to prevent malicious calls. TouchPal employs a reputation-based black-listing approach to prevent malicious calls. TouchPal allows users to label malicious calls. Based on such information, TouchPal will mark a phone number as malicious when TouchPal is confident to do so. We explain the details in the following.

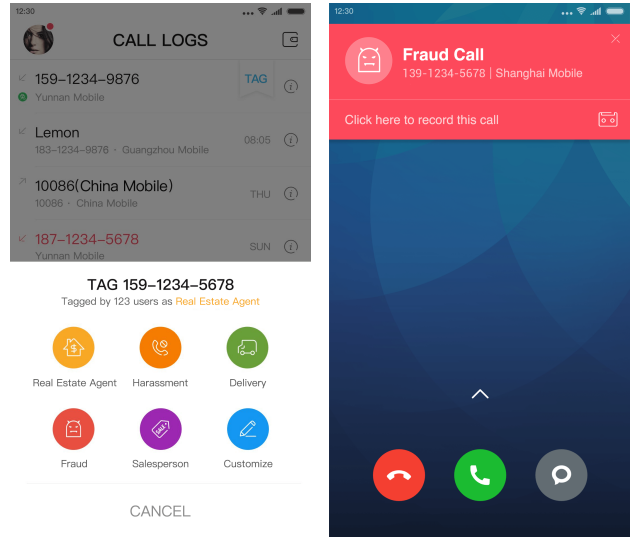

(a) Prompt to label tags  (b) Prompt to reject a scam call

Fig. 1: TouchPal UI on Android

**Information collection.** The user interface for TouchPal to allow users to label malicious callers are provided in Figure 1a. When a call finishes, TouchPal employs some simple heuristics to detect suspicious calls and prompt users with the labeling interface. Doing so can increases the chance that a user may label a call, since most users are reluctant to label calls actively. The heuristics are designed so that the prompt is shown to users only when the call is highly likely to be labeled, so as to reduce the burden to TouchPal users. In particular, if a number is never tagged (almost sure benign) or already blacklisted (almost sure malicious), TouchPal will not prompt its users; only when TouchPal is not sure about a number, it will prompt its users for tagging. Even if the prompt is not showing up, TouchPal still provides a button in the call history to label a malicious call.

A TouchPal user can choose one among five pre-defined tags to label a call. The sixth tag allows users to customize their tags. The five built-in tags are *Real Estate Agency*, *Harassment* (for spams), *Delivery*, *Fraud* (for scams), and *Salesperson*. These categories are created based on the an internal survey about what types of calls that TouchPal users mostly want to block. TouchPal also provides information about the most frequently labeled tag and its frequency. In our analysis, we take *Harassment* and *Fraud* as malicious call tags, and others as benign ones.

**Simple reputation-based black-listing.** TouchPal uses a sophisticated and conservative policy to tag as many phone numbers as possible, while minimizing the amount of wrong tags. Multiple information sources are used to tag a phone number. One of the main sources is the tag provided from users. However, it is very common that users may mislabel some phone numbers. TouchPal imposes a threshold, which may differ from 30 to 100, to be confident about the tag of a phone number. For example, 10086, the service number of China Mobile, is frequently labeled as *Harassment* by users, and its threshold is thus set to be very high. Other information sources are also used to confirm the tag. For example, real estate agency typically provides their numbers online, and TouchPal crawls the websites for those numbers to confirm the tag. Note that the threshold used in TouchPal is not static, and may vary from one number to another.

| Field | Explanation |
|---|---|
| user_id | TouchPal user ID |
| call_type | A binary value indicating if this is an incoming or an outgoing call |
| other_phone | The anonymized phone number in this call other than the TouchPal user |
| other_phone_md5 | A unique encryption for each phone number |
| call_date | The timestamp (in seconds) of the start of the call |
| call_duration | The number of seconds the call lasts |
| call_contact | A binary value indicating if the other number is in the contact of the TouchPal user |
| call_tag | The tag of the call |

TABLE I: The structure of data log records.

| | Oct | Nov | Dec | Total |
|---|---|---|---|---|
| Call records | 3,043 | 2,959 | 3,001 | 9,002 |
| Benign call records | 3,017 | 2,933 | 2,979 | 8,929 |
| Malicious call records | 26 | 25 | 22 | 73 |
| Distinct callers | 256 | 248 | 248 | 447 |
| Distinct callees | 299 | 288 | 287 | 519 |
| Distinct TouchPal users | 24 | 24 | 24 | 35 |
| Distinct malicious call numbers | 0.6 | 0.5 | 0.5 | 0.8 |
| Distinct other numbers | 348 | 338 | 335 | 583 |

TABLE II: Statistics of data log (million) from October to December 2016

Note that the reputation from this black-listing approach also serves the ground-truth to build our machine learning models. Since TouchPal only allows its users to tag malicious calls rather than benign ones, it is hard for malicious users to taint the labels as long as there are enough benign users who label malicious call numbers as malicious.

**Malicious call prevention.** TouchPal allows user to configure the default behavior when the phone number of an incoming call is marked with a tag. For example, the user can choose to hang up a malicious call directly without any notification, or choose to prompt the user. Figure 1b illustrates the interface when an incoming call's phone number is marked as *Fraud*. Note that although TouchPal provides a functionality to record the call content, users have to manually turn it on for each call; also, the recording is only available on users' devices and never uploaded to the server.

The prevention in the current deployment is entirely based on the phone number tag, which can be easily circumvented by using techniques such as caller ID spoofing. We consider this issue as an important future direction.

**Other functionalities.** TouchPal also provides other functionalities such as SMS message prevention. In this work, we focus on the malicious call prevention problem.

## IV. UNDERSTANDING MALICIOUS CALLS IN CHINA

In this section, we investigate the call logs to gain insights on malicious callers' behaviors, and shed light on designing machine learning-based malicious call detection algorithms. In the following, we first present the structure of the call log, and some basic statistics. Then, we study the distribution of TouchPal users, malicious calls, and other numbers along several dimensions: (1) provinces; (2) call time; (3) whether caller is a TouchPal user and/or in the callee's contact; (4) incoming and outgoing call volume; and (5) activeness. For the majority of the analysis, we use call logs spanning three months from October to December 2016. For the liveness analysis, we use all call logs of the entire year of 2016.

### A. Data log description

The structure of each data log record is presented in Table I. When a TouchPal user makes or receives a call, a log record will be generated. The call_type field records whether the user receives the call or makes the call. The other number in the call is anonymized by removing all digits except the first few digits which are similar to the area code in US numbers. These digits contain only the provincial information about the number. The salted MD5 of the entire number is also recorded so that it is possible to distinguish between different numbers for our analysis. In doing so, we break the link from data log to the actual phone numbers to maintain the anonymity. Note that the mapping from a user ID to a phone MD5 is highly confidential. We avoid touching this mapping in all our analysis.

The record contains three types of information about the call: (1) the timestamp, which includes both the date and the time; (2) duration in seconds; and (3) whether the other number is in the contact of the TouchPal user. They can be used for predicting malicious calls. Each call record also contains a call_tag field recording whether the call belongs to one of the 6 categories (i.e., normal or one of the five tags). This field is used as the ground truth of our malicious call prediction task.

In our entire work, we use two other tables which provide more information: (1) the province that each TouchPal user belongs to; and (2) the set of all MD5 hash values of TouchPal users.

We compute the basic statistics of the data log from October to December 2016, and report them in Table II. Note that each call between two TouchPal users generates two records: one for incoming call and another for outgoing call. We can observe over 9 billion call records and over 500 million distinct phone numbers in the period of consideration. The scale of the dataset is sufficiently large for us to draw interesting conclusions. We observe that malicious call records are relatively few with respect to the total records, as expected. There are over 73 million malicious call records with around 800,000 numbers used to make malicious calls. In other words, out of every 100 calls, there will be almost one malicious call, showing that the malicious call problem is severe in China. On average, each malicious call numbers makes 91 malicious calls . Thus identifying a malicious call number earlier (e.g., before it has been used to make 10 malicious calls) can help to prevent a significant amount of malicious calls.

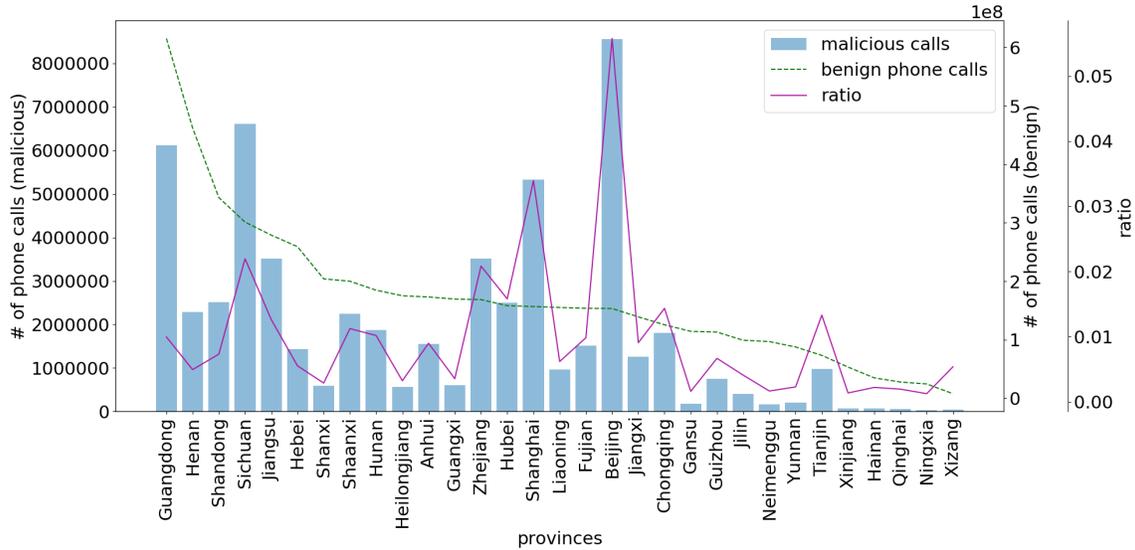

Fig. 2: The histogram of callers from different provinces. The amount of malicious callers, benign callers, and their ratio are computed for each province. The provinces on the x-axis are listed in the descending order of their total amount of calls.

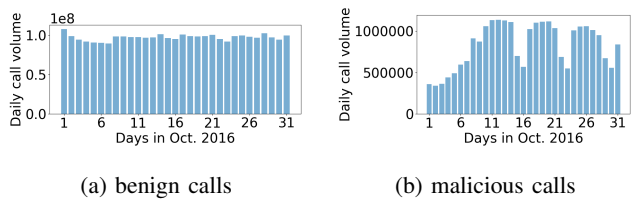

(a) benign calls    (b) malicious calls

Fig. 3: The histogram of the number of benign calls and malicious calls in October 2016. We include call records from all provinces in the results.

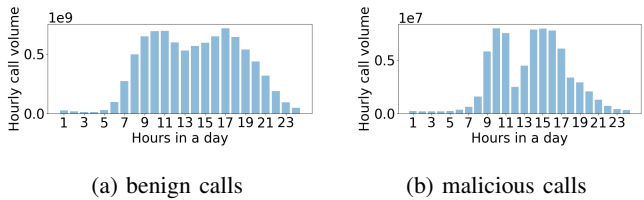

(a) benign calls    (b) malicious calls

Fig. 4: Histogram of hourly distribution for benign call and malicious calls. We include call records from all provinces in the results.

**Validating the ground truth.** In this study, we mainly rely on user taggings to compute the ground truth. To validate whether this is accurate, we randomly sample a subset of malicious call numbers, and call them back. We observe that most of the numbers are not answered albeit multiple attempts in different time. We confirm such numbers as malicious, which constitute the majority. However, there is also a small portion of the numbers that are indeed answered. We find that they belong to personal phone numbers of sales-related professionals (e.g., bankers). Note, although TouchPal provides a specific tag, most Chinese users still consider *cold calls* from sales-related professionals as malicious. So far, we are unable to distinguish such numbers from other malicious calls in a large scale, but we consider them as future work.

**Ethical remarks.** TouchPal users have to agree on the Terms of Use to access to the full functionality of TouchPal. TouchPal notifies its users about data collection through the Terms of Use. Also, TouchPal users have the opt-out option, so that their call history will not be collected, at the cost that the functionality they can use is limited. Our study only touches the users who have agreed to the Terms of Use.

*B. Call distribution across different provinces*

We calculate the histogram of the amounts of malicious call records and all call records, and their ratio for different provinces, and present the results in Figure 2. In the figure, the provinces are listed in descending order based on the total number of calls in the province. We observe that the distribution is very skewed, and a province with a high volume of calls does not imply that it must also have a high volume of malicious calls.

We also make some interesting observations. First, the amount of malicious callers is partially co-related with the Gross Domestic Product (GDP) for each province. For example, among the top-8 provinces with the maximal amounts of malicious callers, 6 of them are also ranked the top-6 based on their GDP in 2016 [23], and the other two are Beijing and Shanghai, i.e., the two largest municipalities in China. This implies that malicious calls in an area may be associated with its economic activities. Second, the amount of total calls may not be associated with the GDP. For example, Shanxi is ranked at 7 based on their total amount of calls, but their GDP ranking is among the bottom 10.

*C. Call distribution across different dates and time*

In this section, we study the distributions of malicious calls and benign calls with respect to (1) dates in a year; (2) days in a week; and (3) hours in a day.

We plot the histogram of benign calls and malicious calls for each day in October 2016 in Figure 3. We observe that the amount of malicious call records during October 1st to October 7th is significantly smaller than other days. This period coincides with the observances of National Day of China when most of the workers were on vacation, and we attribute the observation to this reason. Also, we observe that the malicious call records are reduced significantly three times later, i.e., 15th-16th, 22nd-23rd, and 29th-30th. These three periods are all weekends. We observe similar phenomenon for November and December (see Figure 12 and Figure 13 in the appendix). Therefore, we conclude that the number of malicious calls are associated with whether or not the date of the call is a working day.

On the other hand, however, the correlation between the benign call volume and working days is not as strong as the malicious call volume. We observe a drop of call volume between October 2nd and October 7th, but the amount of benign calls on the 1st is larger than all other days in October. One potential reason may be due to Chinese social convention to make greeting calls at the beginning of a vocation. We lack information to further analyze the reason behind this observation, but this can be of independent interests to some social science disciplines.

We continue to analyze the hourly pattern of benign calls and malicious calls. The histograms are presented in Figure 4. We observe the similar phenomenon: the amount of malicious calls is significantly higher during working hours than off-hours. Also, the malicious calls from noon to 1pm, which is the typical lunch time, are fewer than those during 9am-noon or 1pm-5pm. These observations also confirm that the amount of malicious calls are more correlated with the working hours than benign calls.

Note that our observations are very different than the ones shown in [9], which also aims to understand Chinese malicious calling behaviors. We attribute this to the fact that only less than 700 call records are collected in [9], and thus the results in [9] may not be statistically robust. On the other hand, similar observations have been made based on US's data [15], though some details are different. For example, the hourly call volume histogram from [15] looks more similar to Figure 4a for benign calls than Figure 4b for malicious calls.

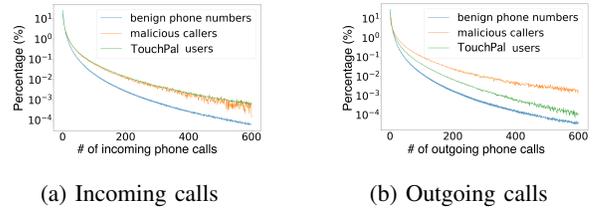

(a) Incoming calls     (b) Outgoing calls

Fig. 5: Distribution of phone numbers based on their incoming and outgoing call volume.

All above observations suggest that the call time may be a useful indicator to distinguish malicious calls from benign ones.

*D. TouchPal users may be unlikely to make malicious calls, but may store malicious callers in their contact*

Now we investigate whether TouchPal users will use their registered phone numbers to make malicious calls. We hypothesize that TouchPal users may unlikely use their registered numbers to make malicious calls because they understand the mechanism how TouchPal prevents malicious calls. We observe that this is indeed the case. Out of the 73 million malicious call records, we identify $103,673$ (i.e., $0.14\%$) records whose callers are TouchPal users. Among these records, we find $2,541$ distinct TouchPal users (i.e., $0.007\%$ of 35 million TouchPal users or $0.32\%$ of 0.8 million malicious callers). Note that our data labeling is not perfect, and it is also possible that some of these numbers are mislabeled. Therefore, we conclude that TouchPal users are unlikely to use their registered phone numbers to make malicious calls.

We further analyze whether a TouchPal user's contact may be a robocaller. Conceptually, the owners of phone numbers stored in a TouchPal user's contact list likely have social relationship with the user, and thus it is unlikely that they are malicious callers. However, we observe the opposite. Among all 73 million malicious call records, we observe 9.9 million of them (i.e., $13.56\%$) whose caller is in the contact list of the callee. We conjecture the reason to be that our malicious dataset may contain personal numbers of sales-related professionals. Some TouchPal users who choose to do business with these professionals may keep their numbers in the contact list, while others may label the numbers as malicious.

*E. Phone number distributions based on incoming call volume and outgoing call volume*

For each phone number, we can compute the volume of its incoming calls. For each $n$, we can compute how many malicious call numbers receive $n$ incoming calls, and then compute their percentage among all malicious call numbers. We plot these percentage values by varying $n$ from $[1, 600]$ as a line in Figure 5a. Similarly, the two other lines are drawn by considering incoming calls of TouchPal users and all benign users. From the figure, we observe a long tail distribution, which is consistent with earlier reports using a smaller dataset [15]. However, we observe that the tails of malicious call numbers and TouchPal users are significantly higher than benign phone numbers. This shows that there is a higher percentage of malicious call users that are making more calls than benign phone users. Therefore, the volume of incoming calls metric of a number in the call log may be an effective indicator for malicious calls.

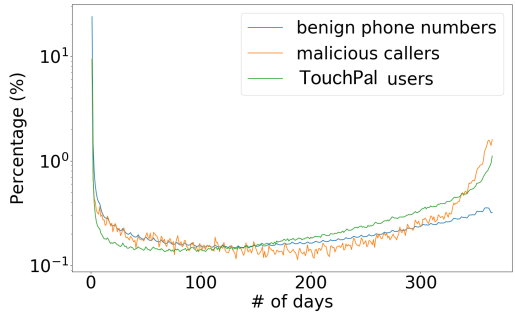

Fig. 6: The distribution of active time

Note that it is non-typical that a non-negligible portion of malicious callers receive a large amount of incoming calls. We conjecture that this is due to our labeled malicious callers may contain sales-related professionals' personal numbers as explained above.

We further create similar plots for outgoing call data in Figure 5b. We observe similar phenomenons: malicious callers tend to make more calls than both TouchPal users and benign phone numbers. Thus this information can be leveraged for malicious call prediction.

### F. Phone number Distribution of active time

We compute the total amount of days that a number is actively used by computing the interval between its first call and its last call in the log. For the analysis to be more informative, we incorporate the data log of the entire 2016. We then compute the distribution in a similar way as for incoming and outgoing call volume analysis, and plot the results in Figure 6.

We observe a "horseshoe"-shaped curve: there is a sharp drop at the beginning, and then gradually grows up at the end. We attribute this to the fact that a user may either stop using a number after a short period of trial, or continue using the number for a long-term. This phenomenon is more significant on the plot for TouchPal users, for whom we have the complete information in the call log.

Surprisingly, there is a much higher percentage of malicious call numbers that are used through the entire year of 2016 than benign numbers or TouchPal users. The bottom of the percentage plot for malicious call numbers appear around 200 days, and its percentage is also lower than both the corresponding values for TouchPal users or benign numbers. This phenomenon is somehow contradicting to the analysis on spammers in other domains, in which a spam account's lifetime is very short. We also conjecture that the reason may be due to our malicious dataset contains not only spam/scam calls, but also sales-related calls. In this case, sales-related numbers will be used for a longer time, since they are some personal phone numbers. We plan to investigate them further in the future.

## V. MALICIOUS CALL PREVENTION USING MACHINE LEARNING

In this section, we present our design of machine learning-based prediction algorithm. In particular, we consider that when a TouchPal user $A$ receives a call from $B$, TouchPal needs to predict whether this will be a malicious call. In the following, we will first discuss the design of features, and then present our choices of machine learning models. Since TouchPal users are unlikely to make malicious calls, in this work, we focus on non-TouchPal users, and leave handling TouchPal malicious callers as a future direction.

### A. Features

As we have discussed earlier, the most challenging problem of malicious call detection is to design a set of informative features. The set of features that we use are listed in Table III.

Intuitively, the basic set of features about a call include (1) whether the caller is in the contact of the callee (is_in_contact); (2) the date and time of the call (weekday and hour); and (3) whether the caller is in the same province as the callee (same_location). However, this basic set of features is not very informative for the detection model to be accurate. We thus extend it in two orthogonal dimensions to extract further information from the call log: *historic information* and *cross-referencing information*. We explain them below.

**Historic information.** The basic features use only information about the current call record. One dimension is to extend to consider all relevant records in the call log. In particular, we retrieve all records involving the current caller, which can be either incoming or outgoing. For each of the records, we can compute a set of features, and thus historic information constitutes a sequence of features vectors.

For basic features, in additional to the basic set of features explained above, there are two additional features that can be computed for each historic record: (1) the call_type indicating whether the record is an incoming call or an outgoing call; and (2) the duration of the call. Note that the call_type feature is excluded from static features, since in the current call, its value is always "incoming call". The duration of the current call is unavailable before the call is answered.

Note that different callers may have different amount of historic records, but a machine learning model typically takes a fixed-length feature vector as input. We will explain how to aggregate all historic feature vectors when discussing the concrete machine learning models.

**Cross-referencing information.** For each call record, either the current one or a historic one, static features considered so far are computed using the information from one record only.

| Category | Feature | Current | Historic | Value | Description |
|---|---|---|---|---|---|
| Static | is_in_contact | ✓ | ✓ | Binary | Whether the other phone number is in the contact of the TouchPal user (call log field) |
| | call_type | | ✓ | Binary | Whether the call is an incoming call or an outgoing call (call log field) |
| | duration | | ✓ | Numeric | The total amount of seconds that a call last (call log field) |
| | weekday | ✓ | ✓ | Numeric | The day in a week when the call starts |
| | hour | ✓ | ✓ | Numeric | The hour in a day when the call starts |
| | same_location | ✓ | ✓ | Binary | Whether the caller and callee are in the same province |
| Cross-referencing | caller_outs | ✓ | ✓ | Numeric | How many outgoing calls that the caller made before the record |
| | caller_ins | ✓ | ✓ | Numeric | How many incoming calls that the caller received before the record |
| | caller_outdegree | ✓ | ✓ | Numeric | How many different phone numbers have been called by the caller before the record in consideration |
| | caller_indegree | ✓ | ✓ | Numeric | How many different phone numbers have called the caller before the record in consideration |
| | callee_outs | ✓ | ✓ | Numeric | The same as caller_outs, caller_ins, caller_outdegree, caller_indegree, but the statistics are computed based on the callee in the considered record rather than the caller |
| | callee_ins | ✓ | ✓ | Numeric | |
| | callee_outdegree | ✓ | ✓ | Numeric | |
| | callee_indegree | ✓ | ✓ | Numeric | |
| | n_call | ✓ | | Numeric | $n$, where the record is the $n$-th call made by the callee in the record |
| | is_redial | ✓ | | Binary | Whether the current caller's last call was with the same number as the current one |
| | gap_to_next | | ✓ | Numeric | The interval (in seconds) between the considered record and the next record made by the same callee |

TABLE III: All 29 input features. For the value type, binary indicates that the feature takes a value from $\{0, 1\}$. Numeric indicates the feature takes an integer value. This value is either used directly as a one-dimensional feature, or converted into a one-hot encoding vector. The vectors for all features are concatenated to form the entire input feature vector.

In addition, we also consider cross-referencing features which are computed by accessing multiple call records in the log.

First, we consider the call log as a stream of records, and thus given any point in the stream, we can take a snapshot to contain only records before the point. In particular, given any record, we can compute the following information of each user based on the record's snapshot: (1) how many incoming calls (ins); (2) how many outgoing calls (outs); (3) how many unique incoming call numbers (indegree); and (4) how many unique outgoing call numbers (outdegree). For each call record (either the current one or a historic one), we can compute these four features for both the caller and the callee in the record. They form the first 8 types of cross-referencing features. As we have observed in Section IV, these statistics are very helpful to distinguish malicious callers from benign callers.

Second, we consider that the current call is the $n$-th call received by the TouchPal user $A$ from the same caller $B$. Then n_call feature takes value $n$. Intuitively, when a TouchPal user $A$ has finished many calls with the caller $B$, it is less likely that $B$ is a malicious caller. Thus, we think n_call can be a useful feature.

Third, we consider whether the caller $B$ just finishes a call with the same TouchPal user $A$ in his last call. This call can be in either direction: from $A$ to $B$ or vice versa. If this is the case, it is likely a redial, and thus less likely a malicious call. We compute it as a binary feature is_redial.

Fourth, the gap_to_next feature considers for each historic record from the callee $B$, the gap (in seconds) between the record and the next one. In fact, we hypothesize that the malicious callers' call pattern may be more prominent than benign users'. Thus it is more likely to identify patterns based on gap_to_next features.

**Remarks.** We want to further remark on the difference between the cross-referencing dimension and the historic dimension. Intuitively, historic features construct *a sequence of feature vectors* as input; and each cross-referencing feature only adds *one additional dimension* to each feature vector. Therefore, these two are orthogonal dimensions of the feature space.

We also remark on the novelty of our features. Previous works have considered several features such as duration [41] and call types and time [20]. However, most of existing works propose ad hoc features grounded on an intuition. In our work, in contrast, the features are designed under the guidance of a large scale measurement study, and proposed systematically following the two general dimensions discussed above. In addition, several features, such as is_in_contact, same_location, and all cross-referencing features are newly proposed in this work.

### B. Machine learning models

The malicious call prediction problem is a standard binary classification problem: classifying an input into either positive (malicious) or negative (benign). We employ several state-of-the-art machine learning models for this problem: neural networks; random forest models [10]; Support Vector Machine (SVM) models [13]; and logistic regression models [14].

We want to emphasize that while non-historic features have fixed dimension, historic ones form a sequence of input vectors. Therefore, we need to convert the sequence into a fixed size input vector. For the above mentioned approaches, we simply take the average of all the vectors in the sequence. In addition, however, we can also employ a *recurrent neural network* [16], which is designed to compute *one embedding vector* from a sequence of input vectors. Due to the space limitation, we explain the model details in the appendix.

We also want to remark that all considered models will emit a score $p$, indicating the probability of the prediction is positive. Therefore, we can set a *model threshold* $\tau$, so that the model will predict malicious when $p \geq \tau$, and benign vice versa. By adjusting $\tau$, a model can make a tradeoff between its precision and recall.

## VI. EVALUATION

In this section, we evaluate different machine learning approaches with respect to their effectiveness to prevent malicious call numbers in comparison with the simple black-listing approach. In the following, we will first explain the experiment setup. Then we will examine different models' performance (1) when trained and tested using data from the same province; (2) when trained on one province and tested on another; and (3) their overall latency. In the next section, we understand the effectiveness of different features by an ablation analysis and examining the features selected by a well-performing model, which is a random forest.

### A. Setup

In this section, we explain the experiment setup. We will begin with implementation details of different models followed by different metrics used to evaluate a model. In the end, we present the details about training and test set construction.

**Model implementation details.** For vanilla neural network, SVM, and logistic regression, we use their built-in implementations from sklearn [26]. We refer to them as NN, SVM, and LR respectively. We implement the LSTM-based RNN model in Tensorflow [5]. We refer to it as RNN. For the random forest models, we use two implementations, one from sklearn, and the other from XGBoost [12]. We refer to these two as RF and XGBoost respectively.

**Evaluation metrics.** We employ two metrics in our evaluation. One is the AUC score, which is a standard metric to evaluate a machine learning model's performance. Note we prefer AUC score over other standard metrics such as precision, recall, and F-1 scores, since AUC score is more robust to the data skew (i.e., in our case, negative examples are $100\times$ more than positive ones).

Note that our desired properties of the model are: (1) most of the benign calls should not be predicted as malicious calls; and (2) a model should identify a new malicious call number by observing the minimal number of malicious calls. To examine how well a model can achieve these two goals compared to the black-listing approach, we design a new metric, **first prediction at label threshold $M$ and precision $p$**, or FP@$(M,p)$ for short. $M$ is the label threshold. That is, a phone number is labeled as a malicious call number once it is labeled at least $M$ times by TouchPal users. In our evaluation, we consider a simplified setting that every malicious call will be labeled so, and thus a black-listing approach will pass at least $M$ malicious calls before the number can be prevented.

We also enforce that the model can achieve a precision $\geq p$ on benign calls. In fact, we can always increase the model threshold $\tau$ to increase the precision of a model. For example, in the extreme case, we can always set $\tau = +\infty$, so that almost all calls are predicted as benign calls to reach a precision of 100%. However, in this case, $\tau$ is set too high to capture any malicious calls. Thus, we define $\tau(p)$ to be the minimal $\tau$ so that the model's precision on the benign calls is at least $p$.

| Model | Beijing | Sichuan | Guangdong | Shanghai | Zhejiang |
|---|---|---|---|---|---|
| RF | **0.9985** | **0.9984** | **0.9978** | **0.9978** | **0.9981** |
| XGBoost | 0.9979 | 0.9981 | 0.9972 | 0.9969 | 0.9977 |
| NN | 0.9978 | 0.9972 | 0.9961 | 0.9966 | 0.9976 |
| RNN | 0.9972 | 0.9962 | 0.9957 | 0.9965 | 0.9975 |
| SVM | 0.9914 | 0.9927 | 0.9895 | 0.9892 | 0.9930 |
| LR | 0.9846 | 0.9822 | 0.9770 | 0.9807 | 0.9848 |

TABLE IV: AUC scores of different models. Each model is trained and tested using data from the same province. The training data uses records between October and November 2016, and the test data uses records in December 2016.

Given a model with $\tau(p)$ and a malicious call number, we are interested in how many call records need be observed before this number can be predicted as malicious. This value is then defined as FP@$(M,p)$. Formally, given a number, whose call records are $R_1, ..., R_n$, FP@$(M,p)$ is defined to be the smallest $i$ such that the model with $\tau(p)$ predicts a input generated from $R_1, ..., R_i$ to be a malicious call. If $i > M$ or none of such $i$ exists, then FP@$(M,p)$ is defined to be $(M+1)$. Given a set of malicious call numbers, we thus can define **averaged FP@$(M,p)$** (or AFP@$(M,p)$ for short) to be the average of all FP@$(M,p)$ values for the malicious call numbers in the set.

We want to comment that in the FP (and AFP) metric, we include the parameter $M$ for soundness. That is, for some cases, the model will never predict a malicious call number as malicious. In this case, its FP value would be $+\infty$ without providing a value $M$, and thus the AFP metric, which would be $+\infty$, is not indicative. We mitigate this issue by including the parameter $M$ in the metric.

**Data construction.** Given a period of time and a province, we construct the training data by selecting all malicious call records from the given province and during the given period. Since there are much more benign call records than malicious call records, we sample a subset of all benign calls which contains the same amount of benign phone numbers as malicious ones to maintain the balance of positive and negative samples in the training set. In each of our experiment, we re-sample the training set 5 times and average their results to make them robust to the sampling procedure.

Similarly, the test set is constructed in the same way. When computing the AFP@$p$ metric, however, we use all benign call records from the given province and in the given period of time. This is because AFP@$p$ is not sensitive to the ratio of positive and negative samples.

### B. Accuracy experiments

In this section, we evaluate different machine learning models in terms of their generalizability. We will first examine the generalizability along the time, and then examine the generalizability to different provinces. We present the details below.

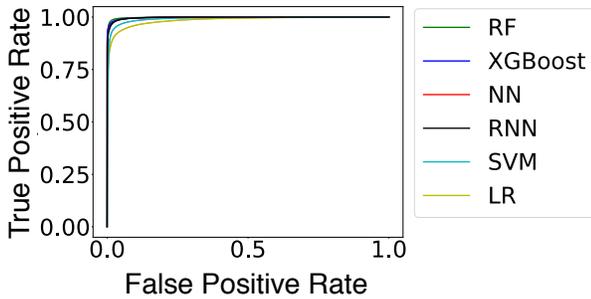

Fig. 7: The ROC curve of different models trained using Beijing's call log records during October and November 2016, and tested on Beijing's call log records during December 2016

*1) Generalizability along the time:* Intuitively, we hope a model trained on the current data logs can serve well in the future. We call this property *the generalizability along the time*. We choose the top-5 provinces with the most amount of malicious calls. For each province, we train a model using the data from this province during October and November, and test it using the data from the same province during December.

**AUC scores.** The AUC results are presented in Table IV. From the table, we can observe almost all methods on any province can achieve an AUC score of 0.985 or higher. This shows that the model is very accurate on predicting malicious calls. Different models are listed in the table in the decreasing order of their performance, from top to bottom. We can observe that for each province, this order is identical to each other. RF (i.e., from the sklearn implementation) achieves the best AUC scores on all provinces, and its AUC performance is at least 0.9978. Also, the XGBoost implementation can achieve similar though slightly lower AUC scores.

The two neural network approaches followed the random forest approaches. Several potential reasons may cause this: (1) the model's capacity is not big enough; (2) the problem has a low-dimension input space, for which a neural network approach may not always be the best; and (3) random forest essentially is an ensemble approach, while we do not use ensemble for our NN approaches. We leave the cause for further investigation. Surprisingly, the RNN's performance is not as good as NN. This may be partialy due to the fact that the performance gain from historic features is not very big. We will further examine this hypothesis in Section VII.

The other two traditional approaches, i.e., SVM and logistic regression, are not as performant as other alternatives. This is reasonable, since both implementations are essentially linear classifiers, which may not be expressive enough to handle the problem.

**ROC curve.** To better understand the AUC scores, we plot the ROC curve for model trained and tested on Beijing's data in Figure 7. From the figure, we can observe that the areas under the curves are almost occupying the entire plot —- and thus the AUC scores are close to 1. This shows that for most models, the threshold $\tau$ can be properly tuned to achieve a very high recall (i.e., the $y$-axis value reaches to 1) while very few benign numbers are predicted as malicious (i.e., the $x$-axis value is close to 0). Therefore, the ROC curve further confirms the effectiveness of our approach.

**Averaged first predictions.** We now present the results using the AFP@$(M, p)$ metric for $M = 10, 20, 30$. The results are presented in Figure 8. From the figure, we observe that XGBoost and NN outperform RF slightly, and the AFP@$(M, p)$ scores of all these three models are always under 5.5. On the other hand, for the other three models, i.e., RNN, SVM, and LR, their AFP@$(M, p)$ scores are close to $M$, when $p$ is set to be large, i.e., 99.99%. The reason is that these models are very hard to achieve a high precision on the benign data to meet the precision requirement, and thus they tend to label any call as benign. In this case, the models are not effective on predicting malicious calls. The random forest models and the non-recurrent neural network do not suffer this issue.

By varying $M$ from 10 to 30, we observe that the AFP@$(M, p)$ score of each of the three best models increases slightly. For example, the score of the best model, XGBoost, raises from 3.57 to 3.90. This is because these models can predict a phone number as malicious call number far earlier before the threshold $M$ is reached.

We consider the total amount of malicious calls that are not blocked by a black-listing approach and our best machine learning approach, XGBoost. Assume there are $N$ different malicious call numbers, then the black-listing approach cannot prevent $N \cdot M$ malicious calls before it starts to be effective; using XGBoost, on the other hand, this amount is $N \cdot (\text{AFP@}(M, p) - 1)$. Therefore, XGBoost can reduce the amount of unprevented malicious calls by $1 - \frac{\text{AFP@}(M,p)-1}{M}$. In our evaluation, XGBoost can achieve a unblocked-call reduction rate from 75.3% (i.e., $M = 10$) to 90.3% (i.e., $M = 30$).

We further investigate the first predictions. In particular, we set each model's $\tau$ to achieve a precision $\geq p$. Then, for each $n \in \{1, ..., 30\}$, we construct the test data by keeping only the first $n$ call records of a number. In this test set, we can compute the *malicious call recall* as the percentage of correctly predicted malicious call numbers using only their first $n$ call records. We call this metric as MR@$(n, p)$, and it provides finer-grained information than AFP@$(M, p)$. By setting $p = 99.99\%$, we plot the MR@$(n, p)$ curve in Figure 9.

We observe that XGBoost and NN can reach a malicious call recall higher than 80% by observing 6 or less calls, but it is hard for any models to reach a malicious call recall higher than 92%. By closely examining the curves, we observe that the NN's recall remains low, and then NN surpasses all other approaches after $n = 7$. XGBoost's recall almost always remains the best until it is surpassed by NN. This shows that NN needs to observe more call records to make effective predictions, while XGBoost requires fewer. We conclude that the best machine learning approaches, XGBoost and NN, can capture most of the malicious calls by observing far fewer call records than a black-listing approach.

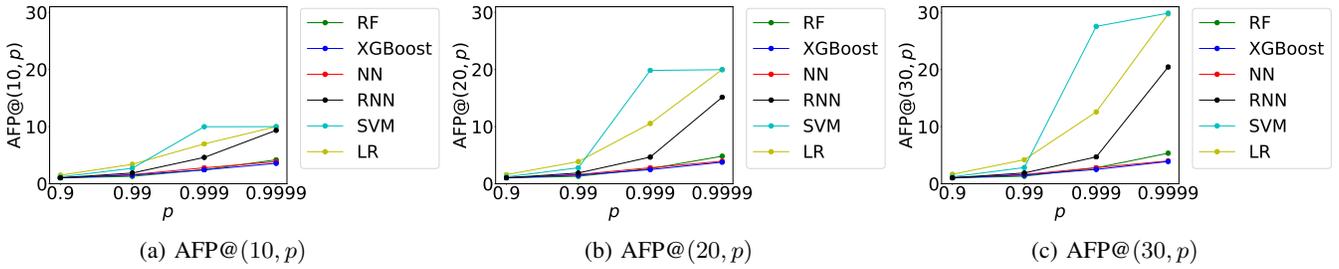

(a) AFP@$(10, p)$  (b) AFP@$(20, p)$  (c) AFP@$(30, p)$

Fig. 8: The AFP@$(M, p)$ of different models trained using Beijing's call log records during October and November 2016, and tested on Beijing's call log records during December 2016.
.

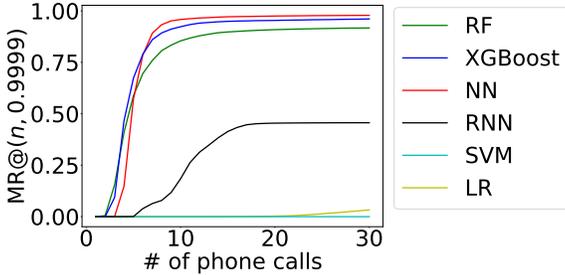

Fig. 9: The MR@$(n, 0.9999)$ of different models trained using Beijing's call log records during October and November 2016, and tested on Beijing's call log records during December 2016

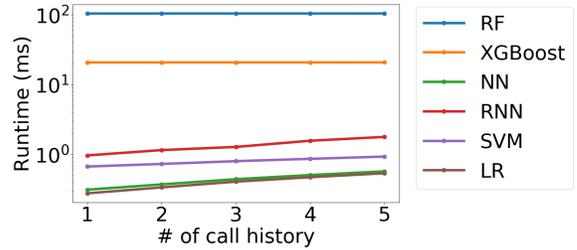

Fig. 10: Runtime of different models

*2) Generalizability to other locations:* In this section, we evaluate whether the model trained on data from one province can generalize to another. We conduct this experiment for two purposes. First, since the sets of phone numbers from different provinces are strictly disjoint from each other, this experiment can give us further insights on whether the model can generalize to unseen models. Second, some area may have too few call records to train an effective model, especially at an early stage of business development in that area. In this case, using a model trained with data from a different province is a promising solution. This experiment is helpful to shed new light on such applications.

We train a model using call logs from Beijing, and evaluate it on data from other provinces. We observe the same phenomenon on the performance of different models in terms of both AUC scores and the AFP@$(M, P)$ values. Due to space limitation, we defer the details of the experiments to the appendix. We also experiment with the models trained using data from different provinces. Our observations are consistent across all experiments. Therefore, we conclude that the model trained on one province with a large call volume can generalize to unseen numbers from another province.

### C. Runtime evaluation

In this subsection, we evaluate the runtime efficiency for different models. The experiments are run on a server equipped with an Intel i7-6900K CPU with 15 cores running at 3.20GHz and 96GB memory. In our evaluation, all data are pre-loaded into the memory to eliminate the I/O latency. We repeat each experiment 5 times and compute the average of different runs as the result.

For each model, we prepare a sequence of historic features and the features for the current record. We compute the *model prediction latency* from processing raw feature sequence till the prediction is made. Note that to retrieve the raw feature sequence, we can leverage the existing key-value store infrastructure deployed in the TouchPal production pipeline, which typically takes 1ms to 10ms to retrieve and update a single record. Thus, an efficient model should not incur a significant overhead on top of it.

For each model, we feed $n = 1, ..., 5$ historic records to construct the inputs to examine the effectiveness of the length of historic features on the model prediction latency. The results are presented in Figure 10. We can observe that random forest models have a high runtime latency ranging from 20ms to over 100ms. This is because the complexity of the model. Note each random forest contains 100 decision trees and each decision tree has three levels. The computation involved in the prediction process of a random forest is much larger than other models.

We observe that for all other models, the runtime latency is less than 2ms, which is a reasonable overhead. In particular, we find that the model NN, which achieves the third best AUC and AFP@$(M, p)$ results, also achieves the second lowest model prediction latency. This is largely due to the simplicity of the model. Therefore, in the scenario when a tradeoff needs to be made between the accuracy and the runtime latency, a non-recurrent neural network may be the best choice in practice.

Further, we observe that when the length of historic feature sequence increases, the model prediction latency for NN,

|       | RF     | XGBoost | NN     | RNN    | SVM    | LR     |
|-------|--------|---------|--------|--------|--------|--------|
| All   | 0.9984 | 0.9979  | 0.9977 | 0.9974 | 0.9913 | 0.9846 |
| -His  | 0.9978 | 0.9978  | 0.9961 | 0.9934 | 0.9890 | 0.9730 |
| -CR   | 0.9444 | 0.9482  | 0.9524 | 0.9556 | 0.9350 | 0.9302 |
| Basic | 0.9079 | 0.9112  | 0.9094 | 0.9084 | 0.9023 | 0.8976 |

TABLE V: Ablation analysis results. "All" indicates all features are used; "-CR" indicates excluding all cross-referencing features from the input; "-His" excluding all historic features from the input; "Basic" is equivalent to "-CR-His". Each cell $(i, j)$ indicates the AUC scores of a model at column $j$ trained with data using input feature set corresponding to the row $i$.

RNN, SVM, LR increases slightly. For the RF and XGBoost implementations, however, the model prediction latencies for different sequence lengths do not exhibit an observable difference. This is because the sequence length only determines the time to construct the input features, while the model prediction latency for random forest models is dominated by the prediction time after the input has been pre-processed.

Note that XGBoost incurs an overhead of 20ms. Therefore, for practical usage, a neural network approach may be more suitable since it achieves comparable effectiveness, but requires much less inference time. However, more optimizations are possible to further accelerate a random forest implementation. We conclude that most machine learning models proposed in Section V incur a reasonably small overhead to be deployed in a real production pipeline.

## VII. UNDERSTANDING THE EFFECTIVENESS OF FEATURES.

In this section, we analyze the effectiveness of the proposed features. We will first perform an ablation study to understand whether adding historic features and/or cross-referencing features indeed helps to improve the performance. We will then analyze one of the most performant decision trees to examine which features are important during the decision making process.

### A. Ablation analysis

In this section, we present an ablation analysis by removing the historic features and/or cross-referencing features. We create training data by using Beijing's October and November call logs, and test data by using Beijing's December call logs. The results are presented in Table V. We observe that the AUC scores of any model using only the basic set of features are very low, i.e., around $0.9$. Thus using more features from the call logs are necessary to achieve a better performance.

We further observe that adding any sets of features always improves the performance, though the improvements are different. We observe that adding cross-referencing features (-His) onto basic features can improve the AUC scores by $10\%$; by adding historic features (-CR), on the other hand, the improvement is only around $4\% - 5\%$.

Using all features is the most accurate approach, but we observe that the improvement from adding historic features on top of cross-referencing features is very small (i.e., $< 0.001$). When we want to achieve better efficiency without sacrificing the accuracy, however, it can be more efficient by using only cross-referencing features.

### B. Understanding the features used by a well-performing random forest

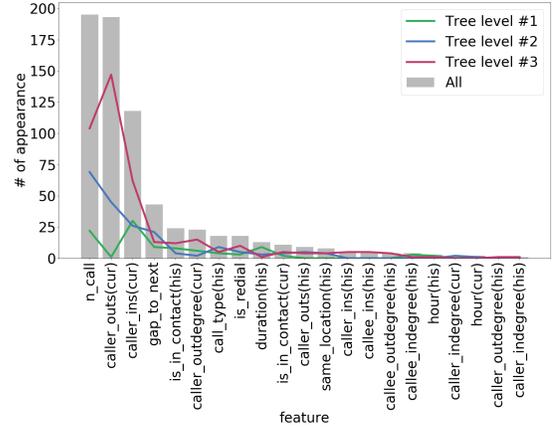

Fig. 11: The histogram of features based on their frequency used in the most performant random forest. The features are listed from left to right in the descending order of their frequencies. We also show three plots based on the frequency of each feature appearing at different level of decision trees in the forest.

We now analyze which of the 29 features are more heavily used by a well-performing model, and examine which features are the most representative to interpret the performance of machine learning models. To do this, we select one the most performant random forest model trained using XGBoost. It contains 100 decision trees with each having 3 layers. We visualize one decision tree in the appendix (see Figure 15).

Each decision tree uses 7 features to reach a leaf (note features used by lower levels may be duplicate). We thus compute the frequency of each feature being used among all 100 decision trees. We plot the histogram based on this frequency in Figure 11. We can observe a long tail distribution. In this model, the top-3 features, namely n_call, caller_outs of the current call, and caller_ins of the current call, are much more frequently used than other features. The most frequently used historic feature in this random forest is is_in_contact, and it is used by less than 25 decision trees. This observation is consistent with our ablation analysis.

Note that features at a higher level will be examined more frequently than the ones at a lower level. We thus analyze the feature distribution at different levels. We plot three lines corresponding to the three levels in Figure 11. We observe that the total frequency of a feature is generally aligned with the each level's frequency. For example, the most frequently used feature in the model, n_call, is frequently used in all three levels. However, there are a few exceptions. For example, the second frequent feature, caller_outs of the current call, is used more on the third level, but less on the top level. Such exceptions are very few.

|        | RF     | XGBoost | NN     | RNN    | SVM    | LR     |
|--------|--------|---------|--------|--------|--------|--------|
| All    | 0.9984 | 0.9979  | 0.9977 | 0.9974 | 0.9913 | 0.9846 |
| Top-10 | 0.9965 | 0.9979  | 0.9967 | 0.9803 | 0.9905 | 0.9784 |

TABLE VI: Different models' AUC scores by using all features and only the top-10 features from Figure 11.

We observe that only 21 out of the 29 features (i.e., 70%) are used in the random forest model. If we cut the long tail by using only features that are used more than 10 times, then only the top-10 features are necessary. We now examine whether this top-10 features used in the random forest is indicative enough for the malicious call prediction problem. In particular, we use the same setup as our ablation analysis, and evaluate different models' performance using only these 10 features. The results are presented in Table VI. We can observe that, the XGBoost implementation's performance does not change at all. This is particularly because the top-10 features are chosen to flavor the XGBoost implementation. We observe that all other models' AUC scores degrades by 0.001 to 0.01. Such a degradation is not significant, and the 3 best approaches' AUC scores are still above 0.9965. Therefore, we conclude that by analyzing a random forest model, we can find the most representative features to interpret the performance of machine learning models.

## VIII. DISCUSSION

One limitation of our work is that it cannot effectively handle caller spoofing. This is a result as we have been focusing on blacklisting approaches to block malicious calls based on the numbers. We consider mitigating this issue as an important future direction.

Also, as mentioned earlier, our system currently cannot distinguish very well between scam or spam callers and sales-related callers. As shown in our study, the active time and whether a number is stored in a TouchPal user's contact may potentially be used as features to make such a distinction. We plan to investigate related issues in the future.

However, our approach is not subject to attackers who may want to white-list a particular malicious call number. As explained in Section III, TouchPal employs a blacklisting mechanism, and thus as long as there are enough benign users tagging a specific malicious call number, it will be labeled as malicious regardless of the effort from the attackers. Therefore, our approach is not subject to *data poisoning attackers*, who try to manipulate training data to make the model predict a malicious call number as benign.

Nevertheless, our machine learning approach may still be subject to two types of machine learning attackers. First, although attackers cannot white-list malicious call numbers, they may use poisoning attacks to black-list benign numbers by setting up a farm to tag benign numbers as malicious. Second, our proposed machine learning approach may be vulnerable to *evasion attackers*, who manipulate the test data during model serving time. In particular, there are several features, such as gap_to_next, that can be intentionally manipulated by the attacker to mimic a benign caller's behavior. We consider mitigating these issues in the future.

## IX. RELATED WORKS

### A. Existing malicious call detection techniques

There have been many prior works discussing malicious call detection, such as white/black-listing [17], [25], [35], [39] and caller's domain reputation [25], [32]. Our work employs machine learning approaches to achieve an accurate solution.

Existing works also design machine learning-based malicious call detection approach, relying on caller behavior [30], [35], recipient behavior [20], [39], social connections [6], [8], [21], [27], [28], and customer feedbacks [17], [18], [32], [37]–[39], [41]. However, all these works assume a server in the telephony network can provide more information about the caller. In contrast, our work is the first machine learning-based approach without relying on any assumptions about the underlying telephony networks.

### B. Telephony malicious call analysis

There have been a variety of systematic studies in malicious call analysis. For example, [15] builds a honeypot with 39,696 phone numbers that are abandoned because former owners received too many unwanted calls. The incoming calls to these phone numbers are treated as malicious calls and analyzed. For more targeted scam calls, such as technical support scams, existing work [22] does a systematic study of both the scams and the call centers behind them. These works typically require recording and analyzing the voice content of the incoming calls, which may break user privacy. Our analysis does not touch users' call content at all, and thus eliminates this privacy concerns.

Several works describe telephone spam ecosystem and provide high level evaluation for the existing techniques [29], [33]. These works highlight the requirements on designing effective malicious call prevention approaches, while our work provides a concrete solution.

## X. CONCLUSION

In this work, we present the first machine learning-based solution without relying on any particular assumptions on the underlying telephony network infrastructures. We propose several techniques to achieve the goal. We first design a TouchPal user interface as a component of a mobile App to allow phone users to label malicious calls. We then conduct a large scale measurement study over three months of call logs, including 9 billion records, and design features based on the results. We extensively evaluate different state-of-the-art machine learning approaches using the proposed 29 features, and the results show that the best approach can reduce up to 90% unpreventable malicious calls while maintaining a precision over 99.99% over benign call traffic. The results also show the models can efficiently make the predictions, and thus can be practically deployed.

## ACKNOWLEDGMENT

We thank the anonymous reviewers and our shepherd Matt Fredrikson for their valuable comments to improve the paper. We thank Xiaojing Liao for helpful discussions. This work was supported in part by FORCES (Foundations Of Resilient CybEr-Physical Systems), which receives support from the National Science Foundation (NSF award numbers CNS-1238959, CNS-1238962, CNS-1239054, CNS-1239166), DARPA under grant no. FA8750-17-2-0091, Berkeley Deep Drive, and Center for Long-Term Cybersecurity, NSFC (61632017, 61772333) and Shanghai Sailing Program (17YF1428200).

Any opinions, findings, and conclusions or recommendations expressed in this material are those of the author(s) and do not necessarily reflect the views of the National Science Foundation.

## APPENDIX

The daily call volume distributions in November and December 2016 are presented in Figure 12 and Figure 13 respectively.

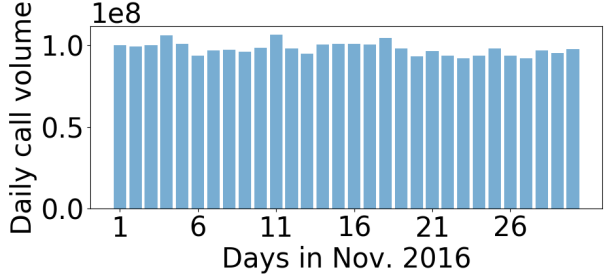

(a) Normal calls

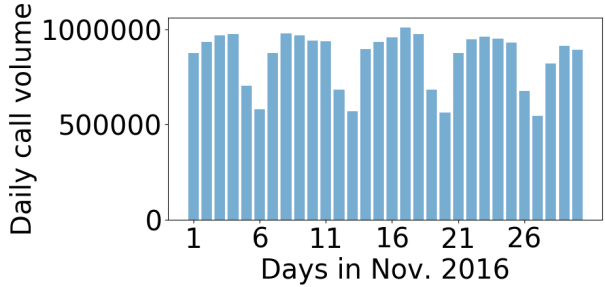

(b) Malicious calls

Fig. 12: Histogram of normal calls and malicious calls in November 2016.

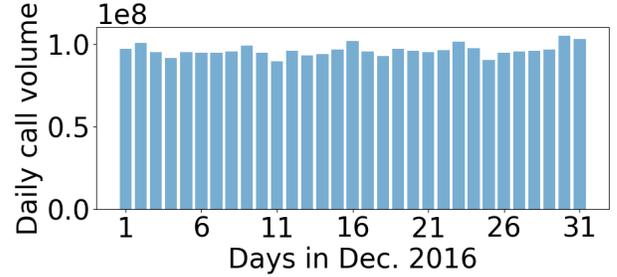

(a) Normal calls

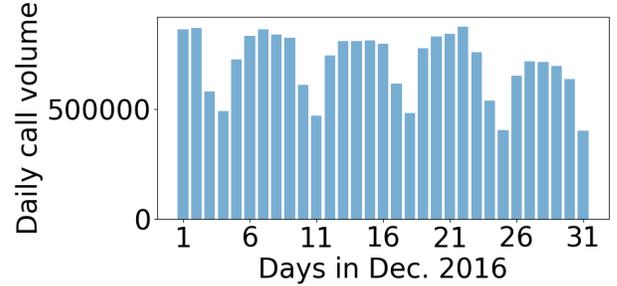

(b) Malicious calls

Fig. 13: Histogram of normal calls and malicious calls in December 2016.

**Explanation to neural network models.**

We start with the simple case when historic features are not used. We employ a two-layer feed-forward network. Note all numeric features are encoded as one-hot vectors. That is, for a feature taking value $v$, assuming the feature's range is $[0, \mathbf{max}]$, then its one hot vector is a $(\mathbf{max}+1)$-dimensional vector, in which the $(v+1)$-th dimension is 1 and all other dimension is 0. One hot encoding the input numeric feature is a common practice when using neural networks.

The hidden states contain 20 neurons. The output, which is a two-dimension vector, is connected with a softmax operator to compute the final prediction. Formally, the prediction can be written as

$$p = \mathbf{softmax}\big(W_1 \times \mathrm{ReLU}(W_2 \times x)\big)$$

where $W_1$ is a $2 \times 20$ matrix, $W_2$ is a $20 \times n$ matrix, ReLU is the standard rectifier function, and $n$ is the input feature dimension. $p$ is a two-dimensional vector, where $p_1$ indicates the probability that the incoming call is a malicious call, and $p_0 + p_1 = 1$ accordingly to the property of $\mathbf{softmax}$. By setting a *model threshold* $\tau$, the machine learning model can predict if the incoming call is a malicious call by checking $p_1 \geq \tau$. By adjusting the model threshold $\tau$, one trained model can make a tradeoff between its precision and recall.

To take the historic features into account, one straightforward way is to treat each record's features as a vector, and compute the average of the feature vectors for all historic records as one fix-length historic feature vector. This historic feature vector is then concatenated with the feature vector for the current call, which becomes the input to the neural network. We refer to this approach as the vanilla NN approach.

However, taking the average may not be the most efficient way to leverage information from the call log. We can consider the historic call records from the callee as a sequence with a variable length. Thus, we can employ a recurrent-neural network (RNN) [16] to convert the sequence into a fix-length embedding. In particular, we employ an LSTM [16] with a hidden state size 16 to compute the embedding, which is then concatenated with the feature vector for the current call. The combined feature is then fed into the neural network above to make the prediction. We refer to this approach as an RNN-based approach.

**Explanation to non-neural network machine learning algorithm.**

Although neural network approaches have achieved significant advancements to handle high-dimensional data, some non-neural network approaches are still more effective in handling low-dimensional inputs. In particular, our inputs include only 29 features, and thus we want to examine whether these non-neural network approaches are more effective than neural network ones. In particular, we are interested in (1) random forest models [10]; (2) Support Vector Machine (SVM) models [13]; and (3) logistic regression models [14]. We briefly explain these models below.

|  |  | RF | XGBoost | NN | RNN | SVM | LR |
|---|---|---|---|---|---|---|---|
| Large Provinces | Guangdong | **0.9979** | 0.9970 | 0.9969 | 0.9961 | 0.9893 | 0.9776 |
|  | Shanghai | **0.9979** | 0.9969 | 0.9969 | 0.9961 | 0.9892 | 0.9793 |
|  | Sichuan | **0.9987** | 0.9983 | 0.9982 | 0.9978 | 0.9926 | 0.9866 |
|  | Zhejiang | **0.9984** | 0.9978 | 0.9976 | 0.9972 | 0.9922 | 0.9847 |
| Small Provinces | Jilin | **0.9973** | 0.9964 | 0.9961 | 0.9955 | 0.9874 | 0.9713 |
|  | Guizhou | **0.9987** | 0.9982 | 0.9981 | 0.9979 | 0.9941 | 0.9865 |
|  | Anhui | **0.9986** | 0.9979 | 0.9978 | 0.9975 | 0.9927 | 0.9845 |

TABLE VII: This table presents the AUC scores of different models when trained on Beijing's call log from October to November, and tested on different provinces' December's call log. "Large provinces" indicate that the top-5 provinces with the largest amount of malicious calls; "Small provinces" indicate the three provinces with the least amount of malicious calls.

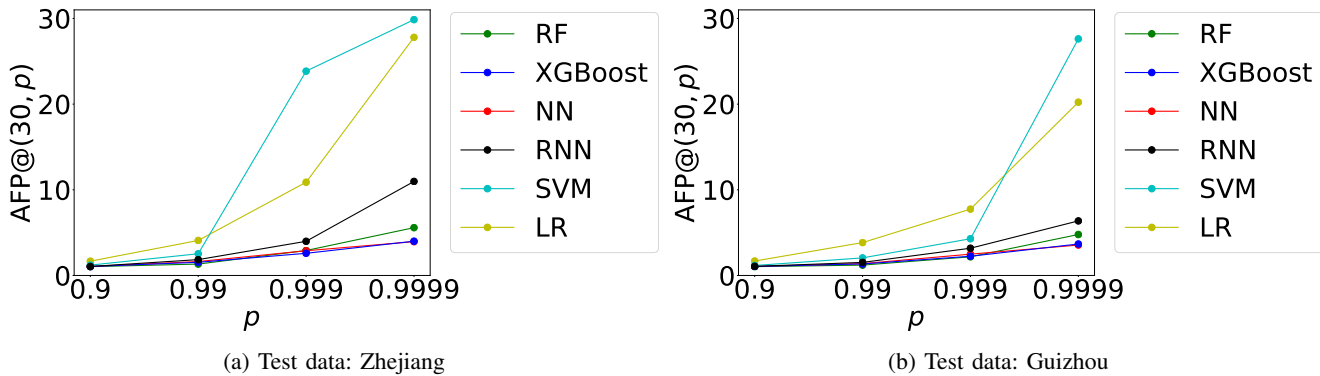

(a) Test data: Zhejiang  (b) Test data: Guizhou

Fig. 14: AFP@$(30, p)$ results for different models trained using Beijing's call logs.

**Random forest models.** A random forest is a collection of decision trees. Each decision tree is a tree whose each internal node labels a feature and a threshold. When making prediction, a decision tree model traverses the tree from the root to a leaf and determines to move left or right depending on whether the value of the input feature labeled on the node is smaller than the threshold or not. Each leaf is associated with a real value, and the value on the leaf at the end of the traversal is returned as its output. A random forest model makes the decision by averaging all values computed from each decision tree in the forest to receive a final value $p$. Again, the decision can be made by setting a threshold $\tau$, in a similar manner as the neural network approaches explained earlier.

**SVM models.** The SVM model is designed to handle the problem when the training data is not linearly separable. In particular, it employs a mapping $\phi$ specified by the user to map the input feature into a high-dimensional space, and then train a model $y = w \cdot \phi(x) + b$, such that the decision plane defined by $w$ and $b$ maximizes the margin, while allowing a few training data to be misclassified. Typically, the $\phi$ is provided as a *kernel function* $\kappa$, such that $\kappa(x, x') = \phi(x) \cdot \phi(x')$. The prediction is also made by using the $\kappa$ function directly. In our case, we employ the linear kernel function to train the SVM model. Note that SVM also emits a real value, which can be used for prediction.

**Logistic models.** The logistic model can be considered as a one-layer neural network: $p = \sigma(wx + b)$, where $\sigma$ is the sigmoid function. This model is commonly applied in industrial applications due to its simplicity and efficiency. However, it may not be as efficient as other alternatives. The output $p$ can be used to make predictions.

Note that all these models take a fix-length input. Thus, we employ the same method in the vanilla NN approach to compute one historic embedding for all historic records.

**Detailed evaluation results for generalizability to other locations.**

In particular, we construct the training data using Beijing's October and November 2016's call logs. We choose 7 other provinces, 4 with large call volumes and 3 with small ones, and use their December 2016's call logs to construct 7 test sets respectively. We train the model using the same training set, and evaluate them on the 7 different test sets respectively. The AUC results are reported in Table VII. We observe that the performance of different models are consistent with previous experiments. Each model trained using Beijing's data can achieve a comparable AUC performance to the model trained using the test province itself, which shows that the model can indeed generalize to unseen numbers from a different location.

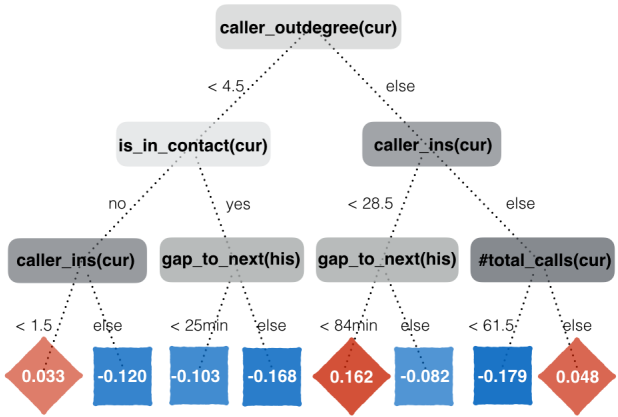

Fig. 15: One example decision tree in the best random forest model trained using XGBoost. Each internal node indicates one feature to be examined. One of its down-going edge is labeled with a checking condition, while the other is labeled with "else". Each leaf node is associated with a value.

By taking a close look, interestingly, we observe that when the model is trained using data from Beijing, its AUC score on another province is even slightly higher than the model trained using the data from the tested province itself. For example, the AUC score of RF on Guangdong is $0.9979$ when trained using Beijing's data, while the value is $0.9978$ when the model is trained using Guangdong's data. Since Beijing has the largest amount of malicious call records, this shows that a larger training set may help to improve the performance.

We present the AFP@$(M, p)$ results for $M = 30$ and the test sets constructed using call log records from Zhejiang (large call volume) and from Guizhou (small call volume) in Figure 14. We make similar observations as previous experiments: (1) the ranks of different models with respect to the AFP@$(M, p)$ are in general consistent with previous observations; (2) random forest models and the NN model's AFP@$p$ values are all below 5.

**Visualization of a decision tree.** In Figure 15, we visualize one decision tree in the random forest trained using XGBoost on October and November's data in Guangzhou.